# Spin Polarization of the 12/5 Fractional Quantum Hall Effect

Chi Zhang[1,2,§], Chao Huan[3,#], J. S. Xia[3], N. S. Sullivan[3], W. Pan[2,*], K.W. Baldwin[1], K. W. West[1], L. N. Pfeiffer[1], and D. C. Tsui[1]

[1]*Department of Electrical Engineering, Princeton University, Princeton, NJ 08544*

[2]*Sandia National Labs, Albuquerque, NM 87185*

[3]*University of Florida and National High Magnetic Field Lab, Gainesville, FL 32611*

Abstract

We have carried out tilt magnetic field (B) studies of the $\nu=12/5$ fractional quantum Hall state in an ultra-high quality GaAs quantum well specimen. Its diagonal magneto-resistance $R_{xx}$ shows a non-monotonic dependence on tilt angle ($\theta$). It first increases sharply with increasing $\theta$, reaches a maximal value of $\sim 70$ $\Omega$ at $\theta \sim 14^o$, and then decreases at higher tilt angles. Correlated with this dependence of $R_{xx}$ on $\theta$, the 12/5 activation energy ($\Delta_{12/5}$) also shows a non-monotonic tilt dependence. $\Delta_{12/5}$ first decreases with increasing $\theta$. Around $\theta = 14^o$, $\Delta_{12/5}$ disappears as $R_{xx}$ becomes non-activated. With further increasing tilt angles, $\Delta_{12/5}$ reemerges and increases with $\theta$. This tilt B dependence at $\nu=12/5$ is strikingly different from that of the well-documented 5/2 state and calls for more investigations on the nature of its ground state.

In a strong magnetic (B) field and at very low temperature (T), a two-dimensional electron system (2DES) can display many novel many-body quantum ground states due to the strong electron-electron (e-e) interactions. A celebrated example is the fractional quantum Hall effect (FQHE). Since the first discovery of the FQHE at the Landau level filling factor $\nu$ = 1/3 [1, 2], many FHQE states have been observed in the lowest Landau level. Here $\nu$=nh/eB, n is the 2DES density, h is the Plank constant and e is electron charge. Remarkably, almost all these FQHE states can be well understood under the non-interacting composite fermion (CF) model [3-5], where the FQHE state of electrons at $\nu$=p/(2mp±1) is mapped onto the integer quantum Hall state of CFs at an effective Landau level filling $\nu^*$ = p, where p is an integer and m = 1,2,3,.. At $\nu$=1/2m, the 2DES forms a Fermi sea of CFs with an effective mass m* which reflects the strong e-e interactions. Both theoretical calculations and experimental measurements have shown that m* $\propto$ $e^2/\varepsilon l_B$, the Coulomb energy in high magnetic fields. Here $\varepsilon$ is the dielectric constant, $l_B$ = $(h/2\pi eB)^{1/2}$ is the magnetic length. An empirical formula, $m^*/m_e \approx 0.26 \times B_\nu^{1/2}$, was deduced for the CFs at $\nu$=1/2 and 1/4 [6-8], where $m_e$ is the free electron mass. Later, in a series of beautiful experiments, it was shown that the CFs also carry a spin, and the effective Landé g-factor g* ~ 0.6 [9,10].

Surprisingly, the non-interacting CF model appears to fail to explain the FQHE in the second Landau level. For example, a pairing mechanism has to be invoked for the so-called 5/2 state, an even-denominator FQHE state [11]. Even for the odd-denominator FQHE states, such as the 7/3, 8/3, and 12/5 states, finite-size numerical calculations have suggested that the correlation in these states are different from that in the lowest Landau level [12-13] and, consequently, they may not be viewed as the integer quantum Hall states of non-interacting CFs. In fact, the 12/5 state, for example, was shown to have a large overlap with the exotic parafermionic state [14] and its ground state is spin polarized.

In recent years, the 12/5 state has attracted growing interest [14-25] due to its superior potential in performing universal topological quantum computation than the 5/2 state [26,27]. On the other hand, in contrast to the well-documented 5/2 state, much less experimental work has been carried out on this state, partially due to its very fragile nature and an extremely small energy gap. Up to date, except for the observation of a well developed quantum Hall plateau at this filling [28,29] there is no direct experimental evidence to support this state being a parafermionic or non-Abelian state.

In this letter, we present our tilt magnetic field dependence results in examining the spin-polarization of the 12/5 state. It was observed that the diagonal magneto-resistance $R_{xx}$ at $\nu=12/5$ shows a non-monotonic dependence on tilt angle $\theta$. It first increases sharply with increasing $\theta$, reaches a maximal value of $\sim 70\ \Omega$ around $\theta \sim 14°$ (at which the total B field $B_{total} = B_{perp}/\cos(\theta)$ = 4.8T.). $R_{xx}$ then decreases with further increase of $\theta$. Correlated with this dependence of $R_{xx}$ on $\theta$ at $\nu=12/5$, the 12/5 activation energy ($\Delta_{12/5}$) also shows a non-monotonic dependence. $\Delta_{12/5}$ first decreases with increasing $\theta$ and vanishes around $\theta = 14°$, when $R_{xx}$ becomes non-activated. With further increasing tilt angles, $R_{xx}$ becomes activated again and $\Delta_{12/5}$ increases with $\theta$. This tilt B dependence of $R_{xx}$ at $\nu=12/5$ and of $\Delta_{12/5}$ are similar to the composite fermion FQHE states at $\nu =$ 2/5 and 8/5 in the lowest Landau level, which now is generally believed to be due to a spin transition. Our results thus call for more investigations on the nature of the 12/5 ground state.

The ultra-high quality 2DES specimen we used in this experiment is a symmetrically doped $Al_{0.24}Ga_{0.76}As/GaAs/Al_{0.24}Ga_{0.76}As$ quantum well (QW). The well width is 30 nm and the set-back distance is 80 nm on both sides of QW. The low temperature 2DES density n = $2.7×10^{11}$ $cm^{-2}$ and mobility µ = $31×10^6$ $cm^2/Vs$ were established after a red light-emitting-diode illumination for several hours at T ~ 4.2K. The size of the sample is about 4 mm × 4 mm with eight indium contacts placed symmetrically around the edges, four at the sample corners and four in the center of the four edges. Our ultralow temperature measurements were conducted in the same setup as in Ref. [30], where the sample can be tilted in-situ by a hydraulic $^3$He rotator. During the tilting process, the perpendicular B field ($B_{perp}$) is fixed for each Landau level filling, while the total B field increases with increasing tilt angle according to $B_{total} = B_{perp}/cos(\theta)$. The in-plane B field is aligned with [110] crystal direction. Standard low-frequency lock-in technique is utilized to measure the diagonal resistance $R_{xx}$ (the excitation current perpendicular to the in-plane B field when under tilt), $R_{yy}$ (current parallel to in-plane B field) and Hall resistance $R_{xy}$. The measurement frequency is ~ 8 Hz and the excitation current is 2-5 nA. During the course of this experiment, the same specimen was thermally recycled from room temperature to the fridge base temperature four times. Data from each cool-down are consistent with each other.

Figure 1 shows $R_{xx}$ and $R_{xy}$ traces taken at the first cool-down at a fridge temperature of ~ 20 mK. In this high quality specimen, well developed FQHE states are observed at ν=14/5(2+4/5), 8/3(2+2/3), 5/2, 7/3(2+1/3), 16/7(2+2/7), and 11/5(2+1/5), evidenced by strong $R_{xx}$ minima and quantized Hall plateaus. Developing FQHE states are also observed at $\nu$=19/7(2+5/7), 12/5(2+2/5), 19/8(2+3/8). The observation of these states is consistent with previous work [28-40]. Besides the above now-generally accepted FQHE states, $R_{xx}$ minima are also observed on both sides of the 5/2 state at B ≈ 4.41T and 4.53T. The one at B ≈ 4.53T can be assigned to the Landau level filling of $\nu$=2+6/13, consistent with a previous study [29]. Surprisingly, in this sample, a quite strong $R_{xx}$ minimum is also observed at $\nu$=21/8(2+5/8), the particle-hole conjugate state of the $\nu$=19/8(2+3/8) state. However, this minimum disappears in the $R_{yy}$ trace. It is not clear at the present time whether this disappearance of $R_{yy}$ is extrinsic (such as due to non-perfect ohmic contacts) or intrinsic (such as due to the formation of an anisotropic phase [41] at this filling). Furthermore, between B = 4.9 and 5.0T, there are three $R_{xx}$ local minima. The two at B = 4.92 and 5.00T correspond to the Landau level fillings $\nu$= 25/11(2+3/11) and 29/13 (2+3/13), respectively. The third minimum at 4.95T, however, is not at any apparent rational filling factor even through it is close to 9/4 (2+1/4). All these three minima disappear in the $R_{yy}$ trace. Further studies are needed to clarify the origins of these new minima.

In the left column of Figure 2, $R_{xx}$ traces are displayed at three selected angles, θ = 0°, 14°, and 27°. The 12/5 state first becomes weaker with increasing tilt angle (shown for θ = 14°), but it becomes a little bit stronger as the tilt angle is further increased (θ = 27°). This trend is corroborated in the $R_{xy}$ plot and the temperature dependence of $R_{xx}$. Shown in the middle column, at θ = 0°, a Hall plateau is clearly visible at $\nu$=12/5. At θ = 14°, the plateau disappears and $R_{xy}$ displays roughly linear B dependence. At the tilt angle is further increased to θ = 22°, a kink starts to develop at ν=12/5. The temperature dependence data is shown in the right column. At θ = 0°, $R_{xx}$ is activated. Though the change of $R_{xx}$ is small over the temperature range, nevertheless, an activation energy $\Delta_{12/5}$ can be obtained from fitting the data according to $R_{xx} \propto \exp(-\Delta_{12/5}/2k_BT)$, and $\Delta_{12/5} \sim 30$ mK. At θ = 14° degrees, the $R_{xx}$ does not show an activated behavior. Instead, $R_{xx}$ decreases with increasing temperature. As θ continues to increase to θ = 27°, $R_{xx}$ becomes activated again, albeit the activation energy at this tilt angle is much smaller, ~ 3mK. Before we discuss Fig. 3, we want to point out that the four huge magnetoresistance peaks associated with the re-entrant integer quantum Hall effect [33] disappear quickly as the specimen is tilted away from the sample normal, a phenomenon first reported in Ref. [30].

In Figure 3a, the $R_{xx}$ and $R_{yy}$ values at ν=12/5 are plotted as a function of $B_{total}$. The data were taken at the base temperature of ~ 10 mK in the second cool-down. Two features need to be emphasized. First, in the studied tilt angle range, the values of $R_{xx}$ and $R_{yy}$ are roughly the same. The tilt induced anisotropic electron transport is not observed. Second, the diagonal magneto-resistance shows non-monotonic tilt B field dependence. It first increases sharply from a value of ~ 7 Ω at θ = 0° to a maximum of ~ 70 Ω at θ ~ 14°. With further increasing θ, $R_{xx}$ and $R_{yy}$ decrease gradually to ~ 35 Ω at θ ~ 40°.

In Figure 3c, the 12/5 activation energy data from two cool-downs are displayed. It also shows a non-monotonic $B_{total}$ dependence. $\Delta_{12/5}$ decreases quickly from ~ 30 mK at θ = 0° to ~ 5 mK at θ = 10°. Between θ ~ 10° and 25° (the gray region), where the curve of $R_{xx}$ ($R_{yy}$) vs. $B_{total}$ displays a broad peak, the magneto-resistance is non-activated. As a result, $\Delta_{12/5}$ can not be deduced. Beyond $\theta = 27°$, $\Delta_{12/5}$ re-emerges and increases with increasing tilt angles.

The tilt angle dependence of the $R_{xx}$ ($R_{yy}$) at $\nu=12/5$ and $\Delta_{12/5}$ is reminiscent of a spin unpolarized ground state under tilt. Indeed, the $B_{total}$ dependence of $R_{xx}$ is very much like that at $\nu=8/5$ [9] and the trace of $\Delta_{12/5}$ versus $B_{total}$ is similar to those of the 8/5 and 2/5 states in the lowest Landau level [42,43]. For both the 2/5 and 8/5 states, the non-monotonic tilt dependence is now generally accepted to be due to a spin transition from a spin unpolarized state to a spin polarized state. In this regard, our tilt magnetic field dependent results indicate a similar spin transition in the 12/5 state.

If this is the case, then, our results apparently are inconsistent with the theoretical models proposed for the 12/5 state being a spin-polarized FQHE state. Rather, they call for the 12/5 state be described as an integer quantum Hall state of non (or weakly) interacting CFs in the second Landau level where it is mapped onto the $\nu^*=2$ state. We show in Fig.3b our fitting according to the model of CFs with a spin. Following the procedure used in Ref. [9], we construct the plot of $B_{total}$ versus $B_{eff}$, where $B_{eff}$ is the effective magnetic field in the second Landau level. For the 12/5 state, $B_{eff} = 5\times(B_{12/5}-B_{5/2}) = 0.93T$. Here, $B_{12/5}$ and $B_{5/2}$ are the perpendicular B field at $\nu=12/5$ and 5/2, respectively. The lines are for $B_{total}/B_{eff} = j\times 2m_e/(g^*m^*)$, where the crossing of CF Landau levels of different spins occurs. j=1,2,… is an integer number. $g^*$ and $m^*$ are the effective g-factor and mass of the CFs in the second Landau level. To our knowledge, neither experimental measurements nor theoretical calculations have been reported on these two parameters. On the other hand, since the effective mass follows an empirical relationship of $m^*/m_e \approx 0.26 \times B_\nu^{1/2}$ for the CFs in the lowest Landau level [6], we assume that this relationship also holds for the CFs in the second Landau level. Consequently, $m^* = 0.55 m_e$ is obtained. For $g^*$, we use the value of 0.6, which has been measured at various even-denominator fillings in the lowest Landau level [10]. With these two values, the lines for j=1 and 2 are drawn in Fig. 3b. It is clearly seen that the peak position in the plot of $R_{xx}$ ($R_{yy}$) versus $B_{total}$ (Fig. 3a) corresponds to the CF Landau level crossing with j=1, just like the 8/5 state in the lowest Landau level [9] and, therefore, strongly supporting the 12/5 state being an IQHE state of CFs.

It is interesting that the non-interacting CF model provides a highly plausible explanation to the tilt B field dependent data in Fig. 2a. On the other hand, the apparent agreement may also signal a new exotic correlated state of composite fermions with spin in the 12/5 FQHE, which allows Zeeman engineering between an unpolarized state (its origin unknown but could be a spin singlet non-Abelian state [15]) and a polarized state (possibly a parafermionic state) [44]. In this regard, we do notice that there are a couple of experimental observations that are inconsistent with the CF model. In general, one expects to see a gap opening right after the LL crossing. This is not observed in Fig. 3c, where there exists a fairly large region where $R_{xx}$ remains non-activated and a true activation gap is not obtainable. Second, the increase of $\Delta_{12/5}$ as a function $B_{total}$ is much weaker after $\Delta_{12/5}$ reappears, when compared to the decreasing rate in the small tilt angle regime. This is different from the 8/5 and 2/5 states, where a similar magnitude was observed before and after the collapse of energy gap [42,43]. At the present time, there is no concrete explanation for this discrepancy. One possibility is that there is no spin transition in the 12/5 FQHE. Instead, the non-monotonic angular dependence could be due to a more complicated mechanism, such as a quantum phase transition from a non-Abelian parafemionic state at zero tilt to an exotic quantum state at high tilt angles. In fact, our very preliminary results show that in the high tilt regime the $\Delta_{12/5}$ value from the temperature dependence of $R_{yy}$ is different from that of $R_{xx}$, suggesting possibly a co-existence phase of the FQHE liquid state with an anisotropic state [45]. Another possibility is that the small slope in the high tilt angle regime is due to that the 12/5 state being in the different electrical subband. In this regard, we have carried out a self-consistent calculation for our sample. It is observed in our tilt range in Fig.3 the 12/5 state (as well as the 5/2 state) remains in the second Landau level of the lowest electrical subband and energetically is far away from the low Landau level of the second electrical subband (however, we note here that the coupling between the electrical bands and magnetic Landau levels under tilt was not considered in the self-consistent calculations.). Moreover, it is clear that in Fig. 4a $\Delta_{5/2}$ decreases continuously with increasing tilt angles and there is no large change in the 5/2 energy gap, as observed by Liu *et al* [40] when the two electrical subbands cross each other.

Finally, it is interesting to observe that the tilt magnetic field induced anisotropic phase was not observed at ν=5/2 in this sample, and $R_{xx}$ and $R_{yy}$ are more or less the same even when the 5/2 FQHE state is destroyed at θ ~ 40°. This isotropic tilt B field dependent behavior at ν=5/2 has also been observed in previous work [46]. The exact origin remains unknown at the present time.

To summarize, we have carried out tilt magnetic field dependent studies of the 12/5 fractional quantum Hall effect state. Its diagonal magneto-resistance $R_{xx}$ shows a non- monotonic dependence on tilt angle, and displays a maximum at θ ~ 14°. We show that this tilt dependence can be understood within the model of CFs with a spin, with appropriate m* and g* values assumed. Furthermore, correlated with the tilt dependence of $R_{xx}$ and $R_{yy}$, the 12/5 activation energy $\Delta_{12/5}$ also shows a non-monotonic B dependence. This tilt B dependence of $R_{xx}$ and $\Delta_{12/5}$ is in striking difference from that of the well-documented 5/2 state and, thus, calls for more investigations of the nature of the 12/5 FQHE.

We would like to thank R.R. Du, J.K. Jain, L.W. Engel, M. Shayegan, N. Read, and E.H. Rezayi for discussions. The experiment was carried out at the high B/T facilities of the NHMFL, which is supported by NSF and by the State of Florida. The work at Sandia was supported by the DOE Office of Basic Energy Science. Sandia National Laboratories is a multi-program laboratory managed and operated by Sandia Corporation, a wholly owned subsidiary of Lockheed Martin Corporation, for the U.S. Department of Energy's National Nuclear Security Administration under contract DE-AC04-94AL85000. The work at Princeton was supported by the DOE under Grant No. DE-FG02-98ER45683, and partially funded by the Gordon and Betty Moore Foundation as well as the National Science Foundation MRSEC Program through the Princeton Center for Complex Materials (DMR-0819860).

§ Present address: International Center for Quantum Materials, Peking University, Beijing, China; chizhang.riceu@gmail.com

# Present address: School of Physics, Georgia Institute of Technology, Atlanta, GA, USA; chao.huan@physics.gatech.edu

* wpan@sandia.gov

Figures:

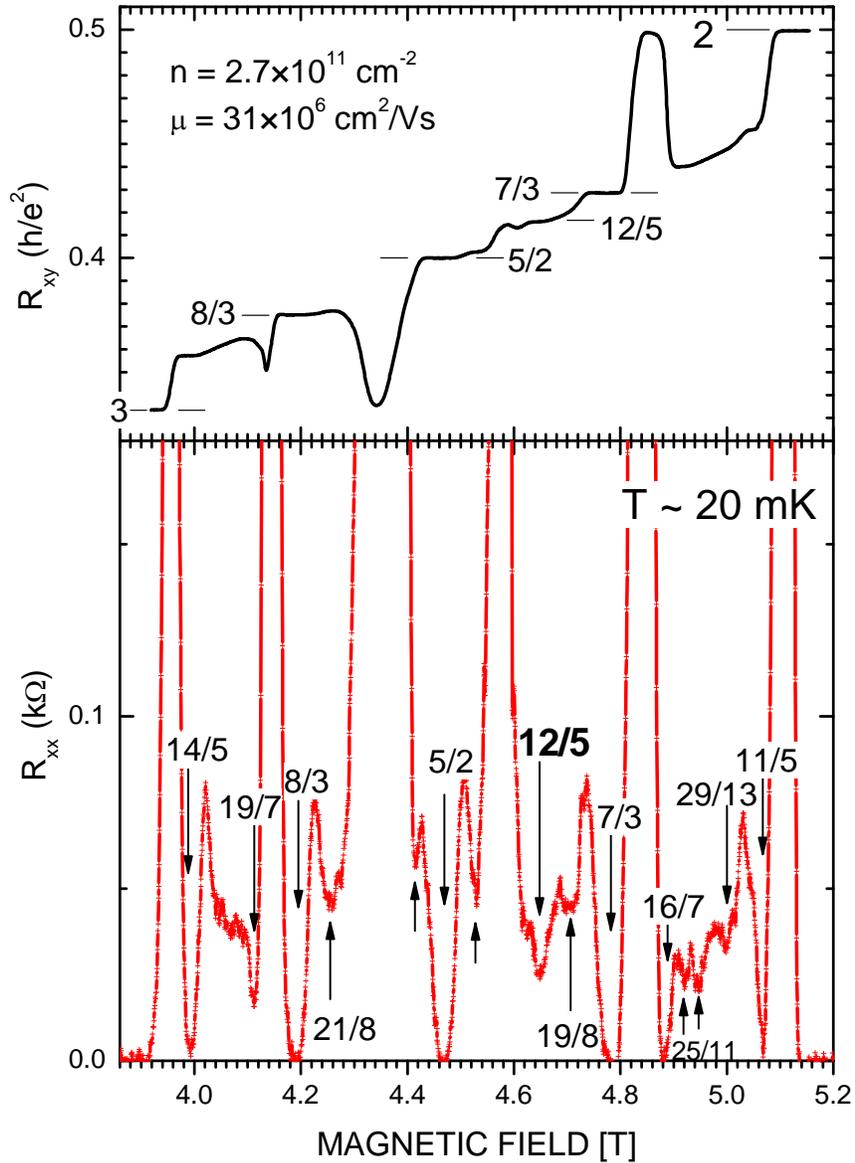

Figure 1 (color online) $R_{xx}$ and $R_{xy}$ traces taken at $\theta = 0°$ under perpendicular magnetic fields. The arrows in the $R_{xx}$ plot mark the fractional quantum Hall states at $\nu$ = 14/5, 19/7, 8/3, 5/2, 32/13, 12/5, 19/8, 7/3, 16/7, and 11/5. Local minima are also observed at $\nu$=21/8, 25/11, and 29/13. The minimum at B = 4.95T is close to $\nu$=9/4. These four minima disappear in the $R_{yy}$ trace. The horizontal lines in the $R_{xy}$ plot show the expected Hall value of each marked QHE state.

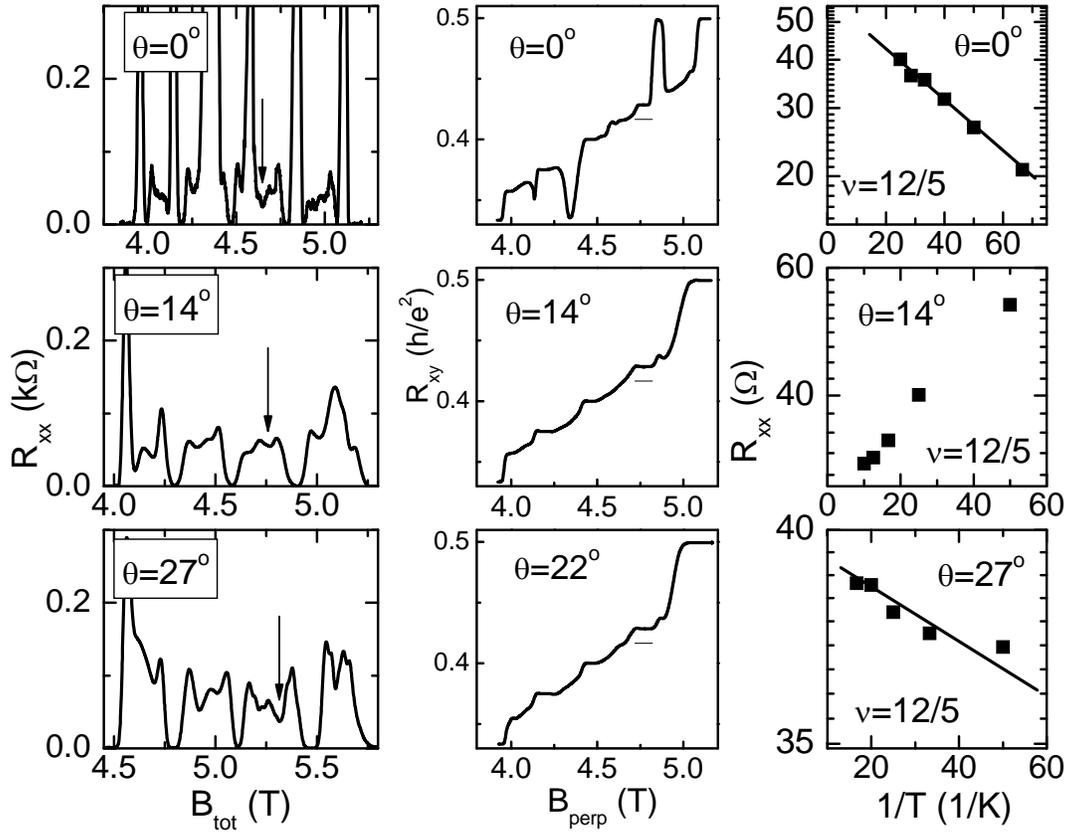

Figure 2 The left column shows the $R_{xx}$ traces at three selected tilt angles, $\theta = 0°, 14°, 27°$. The arrows mark the positions of the 12/5 state. The middle column shows the corresponding $R_{xy}$ traces at $\theta = 0°, 14°,$ and $22°$, respectively. The horizontal lines show the expected Hall value of the 12/5 state. The right column shows in semi-log plot the $R_{xx}$ versus $1/T$ at $\nu=12/5$ at $\theta = 0°, 14°,$ and $27°$. Lines are linear fit.

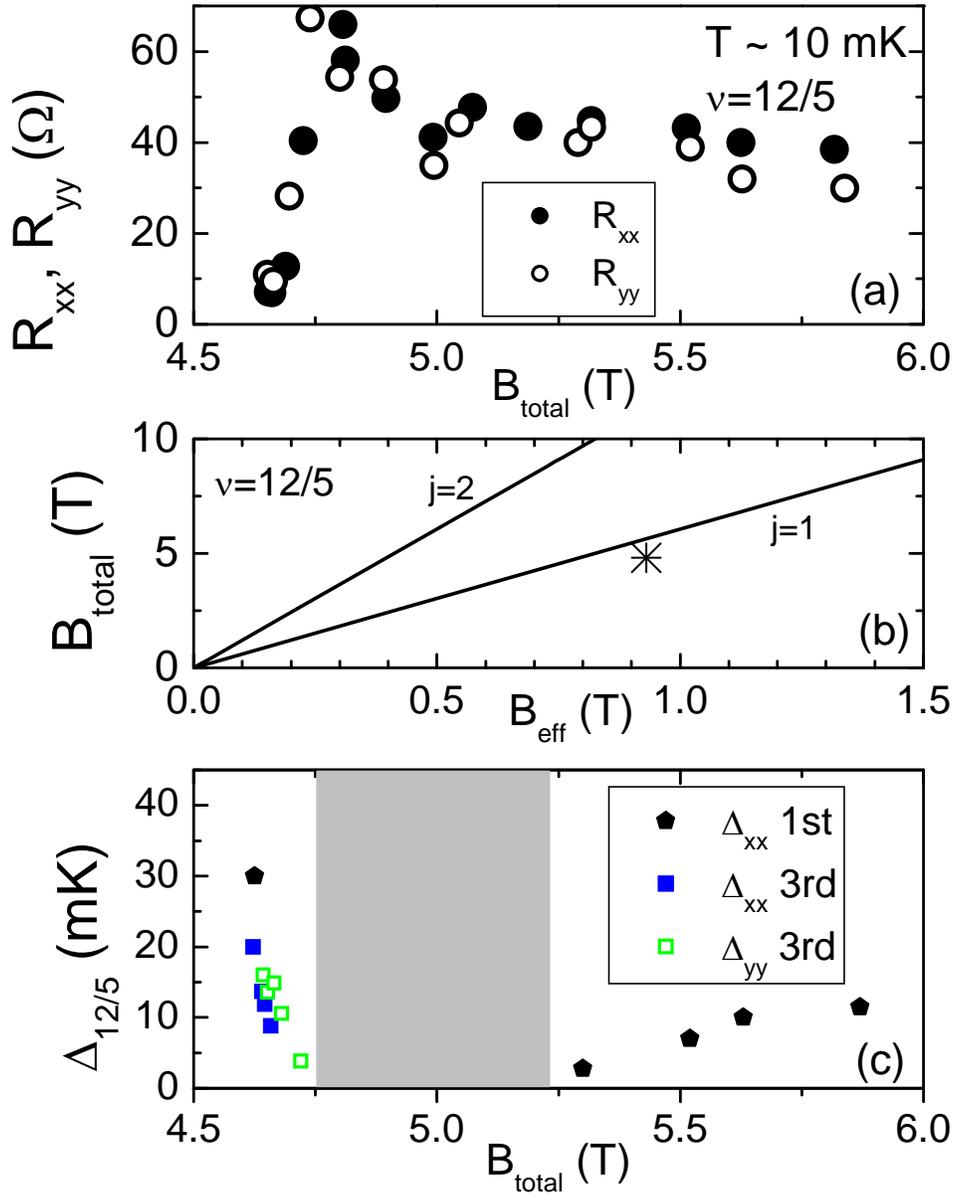

Figure 3 (color online) (a) $R_{xx}$ and $R_{yy}$ at $\nu=12/5$ versus $B_{total}$, measured at $T \sim 10$ mK in the second cool-down. (b) $B_{total}$ versus $B_{eff}$ plot for the maximum at $B_{total} = 4.8$T in (a). The star is the experimental data point. The lines are for $B_{total}/B_{eff} = j \times 2m_e/(g^*m^*)$ with j=1 and 2, respectively. (c) $\Delta_{12/5}$ as a function of $B_{total}$ from two cool-downs. For the third cool-down, $\Delta_{12/5}$ from the temperature dependence of both $R_{xx}$ and $R_{yy}$ is shown. In the gray region, magneto-resistance is not activated and $\Delta_{12/5}$ is not obtainable.

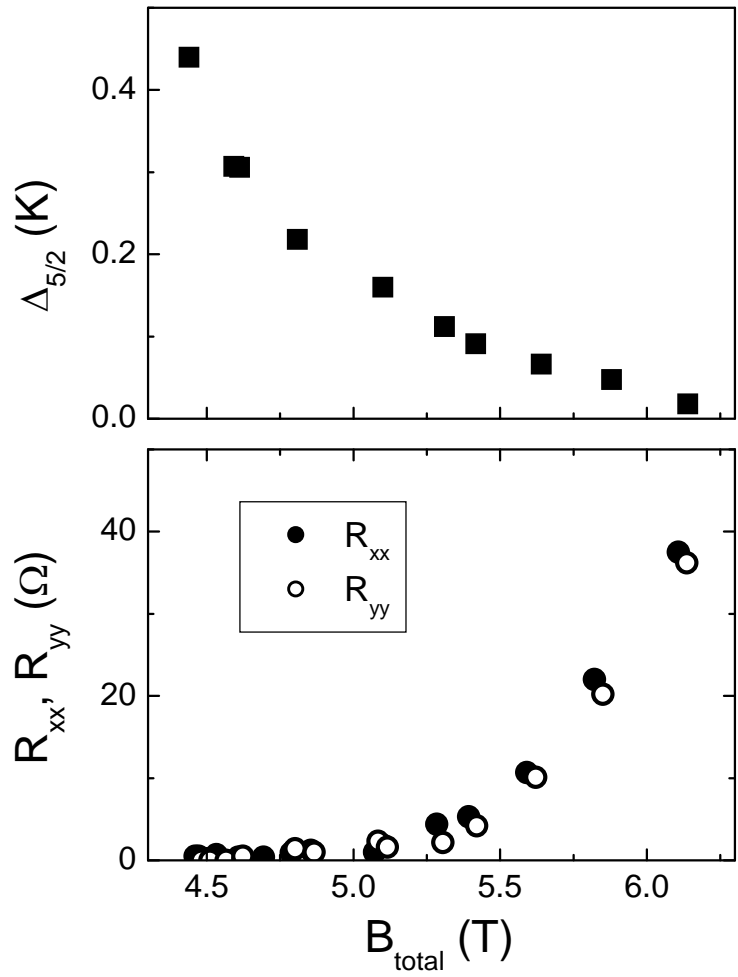

Figure 4 (a) The energy gap of the 5/2 state as a function of $B_{total}$. (b) $R_{xx}$ and $R_{yy}$ at $\nu=5/2$ versus $B_{total}$, measured at T ~ 10 mK.